\documentclass[aps,prl,reprint,groupedaddress,showpacs]{revtex4-1}
\usepackage{graphicx}
\usepackage{amsmath}
\usepackage{hyperref}

\begin{document}

\title{Comment on ``Fast and Accurate Modeling of Molecular Atomization Energies with Machine Learning''}
\author{Jonathan E. Moussa}
\email[]{godotalgorithm@gmail.com}
\affiliation{Sandia National Laboratories, Albuquerque, NM 87185, USA}
\date{\today}

\pacs{82.20.Wt, 31.15.B-, 31.15.E-, 82.37.-j}

\maketitle

\begin{figure}
\includegraphics[width=85mm]{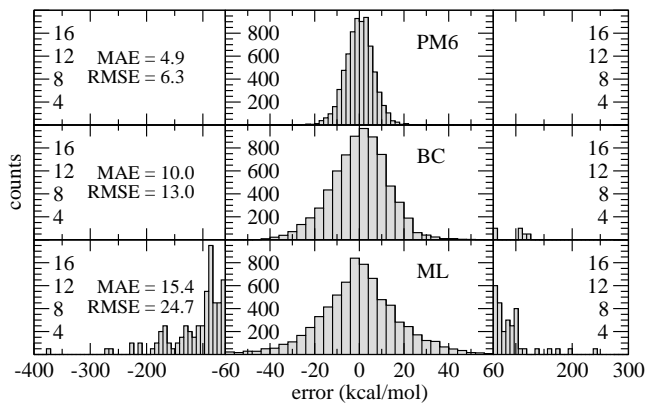}
\caption{\label{error} Error histograms ($E_{\textrm{DFT}}-E_{\textrm{model}}$), mean absolute errors (MAE), and root-mean-square errors (RMSE) for PM6, BC, and ML models compared to DFT
 on the 7169 molecules of the GDB-13 set \cite{GDB} with the formulae C$_v$H$_w$N$_x$O$_y$S$_z$ for $3\le v+x+y+z\le 7$. }
\end{figure}

In a recent Letter \cite{Lilienfeld}, the authors construct a machine learning (ML) model of molecular atomization energies,
 which they compare to bond counting (BC) and the PM6 semiempirical method \cite{PM6}.
However, their ML model was trained and tested on density functional theory (DFT) \textit{energies}
 while BC and PM6 are fit to \textit{standard enthalpies}.
For fair comparison, bond energies are refit to DFT data and PM6 is converted to an electronic energy using per-atom corrections 
\footnote{DFT energies are recomputed using the 6-311$G(3df,2p)$ basis set and PBE0 functional.
The ML model is trained on the 5000 Coulomb matrices of ``model 1k'' \cite{Lilienfeld} and tuned ($\sqrt{\lambda} = 0.0021$ and $\sigma = 24.0$) to minimize test set MAE. Prescribed PM6 corrections \cite{PM6} and bond energies are least-squares fit to the test set. Geometries and bond orders are determined by OpenBabel 2.3.1 \cite{OpenBabel}}.
BC and PM6 both perform better than the ML model and are free of large outliers in their error distributions as shown in Fig. \ref{error}.

\begin{figure}
\includegraphics[width=85mm]{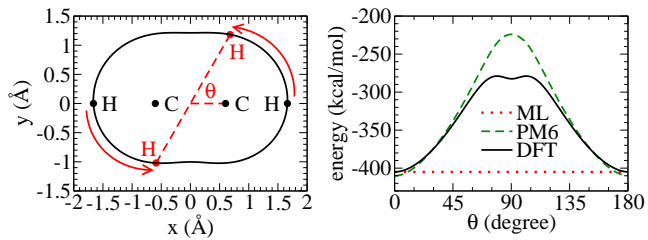}
\caption{\label{path} (color online) A continuous deformation of acetylene.
(left) Hydrogen atoms follow the closed curve with the line connecting them fixed to the origin.
Carbon atoms remain near their equilibrium positions.
(right) Atomization energy as a function of H-origin-C angle. }
\end{figure}

As noted in Footnote 25 of the Letter, some ML model error may originate from the coordinate system choice.
The $n$ eigenvalues of the Coulomb matrix correspond to an equienergy $2n$-dimensional space of $n$-atom molecules rather than one molecule.
For $n = 3$, this corresponds to the $3$ translations and $3$ rotations that naturally preserve the energy of an isolated molecule.
For $n > 3$, the space includes unphysical molecular deformations that destroy structural rigidity.
Fig. \ref{path} shows this with a distortion of acetylene (C$_2$H$_2$) that preserves its ML energy and coordinate, $(53.058,21.149,0.290,0.219)$.

It is suggested in Footnote 25 of the Letter that the $n^2$ sorted entries of a Coulomb matrix
 might be utilized instead of its $n$ eigenvalues as a ML coordinate system.
This eliminates the dimensional deficiency, but produces identical coordinates for homometric molecules \cite{homometric}
 that do not necessarily have equal energies.
 A computationally expensive alternative is the equivalence class of permuted Coulomb matrices with distance metric
\begin{equation}
 d(\mathbf{M},\mathbf{M}') = \min_{\mathbf{P}} \| \mathbf{M}- \mathbf{P}^T \mathbf{M}' \mathbf{P} \|_F
\end{equation}
 for Coulomb matrices $\mathbf{M}$ and $\mathbf{M}'$, permutation matrices $\mathbf{P}$, and the Frobenius matrix norm.

Another possible source of ML model error is its lack of size-consistency.
Even if the energy of two molecules $A$ and $B$ are accurately modeled in isolation,
 there are no guarantees that the well-separated pair of molecules $A+B$ will be similarly accurate.
This requires explicitly filling the chemical compound space with a sufficiently dense set of training molecules,
 which likely leads to an O$(\alpha^n)$ computational complexity for $n$ atoms ($\alpha > 1$).
While benchmarks are favorable for $n \le 7$, the ML model cannot scale favorably compared to
 O$(n)$ classical force fields or O$(n^3)$ DFT/semiempirical methods.
Alternative ML methods \cite{GAP} enforce size-consistency by modeling
 an intensive quantity, per-atom energy, rather than directly modeling the extensive total energy
 and control costs by exploiting nearsightedness \cite{nearsightedness}.
 
\begin{acknowledgments}
Sandia National Laboratories is a multi-program laboratory managed and operated by Sandia Corporation,
a wholly owned subsidiary of Lockheed Martin Corporation, for the U.S. Department of Energy’s
National Nuclear Security Administration under contract DE-AC04-94AL85000.
\end{acknowledgments}

\bibliography{comment}

\end{document}